\documentclass[5p,times]{elsarticle}

\usepackage{algorithm}
\usepackage{algpseudocode}
\usepackage{amsmath}
\usepackage{booktabs}
\usepackage{caption}
\usepackage{color}
\usepackage{xcolor}
\usepackage{colortbl}
\usepackage{float}
\usepackage{graphicx}
\usepackage[hidelinks]{hyperref}
\usepackage{cleveref}
\usepackage{listings} 
\usepackage{makeidx}
\usepackage{multirow}
\usepackage{paralist}
\usepackage[raggedrightboxes]{ragged2e}
\usepackage{rotating}
\usepackage{subcaption}                    
\usepackage{tabularx}
\usepackage{todonotes}
\usepackage[normalem]{ulem}
\usepackage{varwidth}
\usepackage{xspace}
\usepackage{comment}
\usepackage{balance}

\AtBeginDocument{
\definecolor{myorange}{RGB}{255, 243, 219}
\definecolor{myorange2}{RGB}{255, 170, 23}
}

\lstdefinestyle{mystyle}{
    backgroundcolor=\color{gray!10},  
    commentstyle=\color{blue},       
    keywordstyle=\color{red},        
    numberstyle=\tiny\color{gray},    
    stringstyle=\color{cyan},          
    basicstyle=\ttfamily,             
    breaklines=true,                  
    numbers=left,                     
    stepnumber=1,                      
    frame=single                      
}
\journal{Future Generation Computer Systems}

\newcommand\command{\begingroup \urlstyle{tt}\Url}

\begin{document}
\begin{frontmatter}

\title{
Bridging Paradigms: Designing for HPC-Quantum Convergence
}

\author[nccs]{Amir Shehata\corref{cor1}}
\ead{shehataa@ornl.gov}

\author[nccs]{Peter Groszkowski}
\ead{groszkowskip@ornl.gov}

\author[csm]{Thomas Naughton}
\ead{naughtont@ornl.gov}

\author[nccs]{Muralikrishnan Gopalakrishnan Meena}
\ead{gopalakrishm@ornl.gov}

\author[csm]{Elaine Wong}
\ead{wongey@ornl.gov}

\author[cse]{Daniel Claudino}
\ead{claudinodc@ornl.gov}

\author[nccs]{Rafael Ferreira da Silva}
\ead{silvarf@ornl.gov}

\author[nccs]{Thomas Beck}
\ead{becktl@ornl.gov}

\address[nccs]{
    National Center for Computational Sciences, 
    Oak Ridge National Laboratory,
    Oak Ridge, TN, USA
}

\address[csm]{
    Computer Science and Mathematics, 
    Oak Ridge National Laboratory,
    Oak Ridge, TN, USA
}

\address[cse]{
    Computational Sciences and Engineering, 
    Oak Ridge National Laboratory,
    Oak Ridge, TN, USA
}
    
\cortext[cor1]{Corresponding address: National Center for Computational Sciences, Oak Ridge National Laboratory, Oak Ridge, TN, 37831, USA}

\cortext[cor2]{\scriptsize This manuscript has been authored by UT-Battelle, LLC, under contract DE-AC05-00OR22725 with the US Department of Energy (DOE). The publisher acknowledges the US government license to provide public access under the DOE Public Access Plan (\url{http://energy.gov/downloads/doe-public-access-plan}).}

\begin{abstract}
This paper presents a comprehensive software stack architecture for integrating quantum computing (QC) capabilities with High-Performance Computing (HPC) environments. While quantum computers show promise as specialized accelerators for scientific computing, their effective integration with classical HPC systems presents significant technical challenges. We propose a hardware-agnostic software framework that supports both current noisy intermediate-scale quantum devices and future fault-tolerant quantum computers, while maintaining compatibility with existing HPC workflows. The architecture includes a quantum gateway interface, standardized APIs for resource management, and robust scheduling mechanisms to handle both simultaneous and interleaved quantum-classical workloads. Key innovations include: (1)~a unified resource management system that efficiently coordinates quantum and classical resources, (2)~a flexible quantum programming interface that abstracts hardware-specific details, (3)~A Quantum Platform Manager API that simplifies the integration of various quantum hardware systems, and (4)~a comprehensive tool chain for quantum circuit optimization and execution. We demonstrate our architecture through implementation of quantum-classical algorithms, including the variational quantum linear solver, showcasing the framework's ability to handle complex hybrid workflows while maximizing resource utilization. This work provides a foundational blueprint for integrating QC capabilities into existing HPC infrastructures, addressing critical challenges in resource management, job scheduling, and efficient data movement between classical and quantum resources.
\end{abstract}

\begin{keyword}
Quantum Computing, High-Performance Computing, System Integration, Quantum Applications.
\end{keyword}

\end{frontmatter}

 Sections
\section{Introduction}
\label{sec:introduction}

\nocite{*}

\begin{figure*}[!htb]
  \centering
  \includegraphics[width=0.95\linewidth]{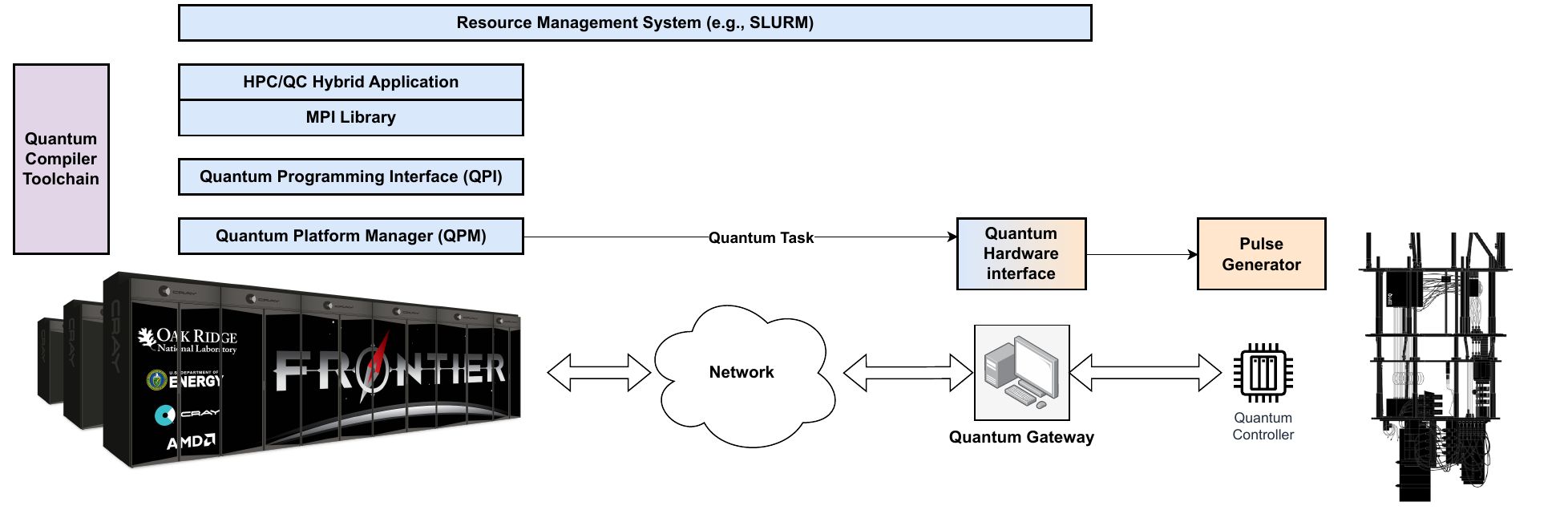}
  \caption{A high-level physical view of QC/HPC integration, overlaid with a simplified representation of the quantum software stack. The quantum gateway can interface with multiple quantum hardware systems of the same type. Additionally, multiple gateways, including virtualized ones, can be configured to communicate with different types of hardware. This separation is designed to minimize workload on the gateway, reducing congestion and preventing over-subscription.}
  \label{fig:highh-level}
\end{figure*}

Quantum Computing (QC) holds immense promise for accelerating scientific discovery across multiple domains, including quantum chemistry, materials science, optimization, and artificial intelligence (AI). However, QC is not expected to replace classical computing but rather serve as an accelerator for tasks best suited to its capabilities, necessitating their integration with their classical counterparts, particularly High Performance Computing (HPC). Similar to how GPUs enhance classical computing through heterogeneous integration, quantum processing units (QPUs) could accelerate specific quantum advantaged algorithms while leveraging classical systems for other computations within larger scientific workflows.

Currently, several major research centers worldwide are actively pursuing QC/HPC integration initiatives. The European Union's EuroHPC project has designated multiple centers for quantum computer installation, including facilities in Czechia, Finland, France, Germany, Italy, Poland, and Spain. Similarly, significant efforts are underway at various national research laboratories in the United States. These initiatives primarily focus on developing optimized software stacks and workflows for various quantum technologies, preparing for the eventual integration of fault-tolerant quantum computers (FTQC).

The Oak Ridge Leadership Computing Facility (OLCF) at Oak Ridge National Laboratory (ORNL) has established considerable expertise in deploying world-leading supercomputers that utilize accelerator technologies, as demonstrated by systems like Frontier and Summit. Through the Quantum Computing User Program (QCUP), ORNL has gained valuable experience in providing access to diverse quantum resources and managing hybrid quantum-classical workflows. In our previous work~\cite{beck2024quantum}, we outlined a comprehensive framework for integrating QC capabilities into HPC environments and detailed the infrastructure requirements for supporting quantum-classical workflows. This experience, combined with ORNL's established processes for supercomputer lifecycle management, provides a strong foundation for developing comprehensive QC/HPC integration strategies.

Building upon our previous integration framework~\cite{beck2024quantum}, this paper advances the state of the art by presenting a detailed software stack design and implementation strategies for quantum-HPC integration. While our previous work established the foundational requirements, this paper focuses on the practical software architecture and design aspects, as well as validation of the software stack in real-world scenarios. Our key contributions include:

\begin{compactenum}
    \item A hardware-agnostic software stack design that supports both current noisy intermediate-scale quantum-era (NISQ) devices and future fault-tolerant quantum computers, while maintaining compatibility with existing HPC workflows and resource management systems.

    \item A comprehensive software stack architecture that addresses critical challenges in quantum-HPC integration, including resource scheduling, job management, and efficient data movement between classical and quantum resources.

    \item An analysis of the proposed software architecture, including its interaction with scientific applications and its integration with external software components such as workflow management systems.

    \item Implementation strategies for supporting both on-premises quantum hardware and cloud-based quantum resources, enabling flexible deployment options for HPC centers. 
\end{compactenum}

\medskip

\section{Framework Overview}
\label{sec:connectivity}

A high-level physical view of the
integration of an on-premises quantum machine is illustrated in Figure~\ref{fig:highh-level}. In the figure, the
quantum machine is connected to a quantum controller. The quantum controller
acts as the interface between classical control hardware and quantum
processors (qubits). The main function of the controller is to generate, manipulate, and read
out signals that control the quantum computation.

A classical compute node is directly connected to the quantum controller and
acts as a ``quantum gateway'' for access to the quantum resource.  The
quantum gateway computer runs classical services, which include resource
management software to efficiently
allocate/reserve the quantum hardware.  The gateway computer also provides
dedicated, low-latency computational resources for any classical operations
that are needed by the quantum controller to satisfy timing constraints. Lastly, the gateway can
host portions of the resource management plugins that are hardware-aware and
used to interface resources elsewhere on the network (e.g., \emph{Frontier}
HPC supercomputer).

The bulk of the QC/HPC software stack resides on traditional classical
HPC resources, e.g., service nodes, compute nodes, etc.  The resource
manager (e.g., SLURM~\cite{schedmd:slurm}) runs on the HPC service nodes and
the compute node where application processes are launched.  The hybrid
application spans the HPC and QC resources.  Figure~\ref{fig:highh-level}
overlays a notional mapping of where the different services and capabilities
reside.  The Quantum Programming Interface~(QPI) is the application facing
layer that is used by the hybrid application on the HPC compute nodes to
interact with the quantum resources.  The Quantum Platform Manager~(QPM) is
the hardware facing layer that provides access to underlying quantum
resources from the HPC compute nodes.  These software layers are described
in more detail in Section~\ref{sec:hpcqc-arch}.

\section{Traditional HPC Accelerators: GPU Example}
\label{sec:back-traditional}

For the foreseeable future, QC will likely be regarded as a
specialized computational accelerator for tasks particularly suited to this
paradigm. Therefore, it is instructive to analyze the integration of GPUs in
classical computing to draw parallels and lessons from GPU/CPU integration. ORNL's Frontier~\cite{10.1145/3581784.3607089} supercomputer utilizes AMD GPUs and has further emphasized the need for specialized accelerators in computer architecture. 

\subsection{GPU Control Interfaces}

Developing efficient GPU programs requires mastering two distinct phases: (1)~Developing and compiling GPU-enabled code, and (2)~Running the resulting binary on the target system. Typically, 
GPU kernels are annotated with specific syntax, such
as \verb|__global__|, to signify that the function is intended for execution
on the GPU. GPU APIs, like CUDA~\cite{cuda:toolkit,cuda2008acmq} or HIP~\cite{amd:hip}, provide a method to interact with
the GPU and can be categorized into several groups. For example, Initialization and Version Management APIs manage the setup and version checking of the GPU environment. Device Management APIs handle device selection, querying device attributes, and setting various device configuration. Execution Control APIs govern how kernels are launched and executed on the GPU. Memory Management APIs handle GPU memory management, including allocation, deallocation, and copying. These APIs are employed within host code, running on the CPU, to set up the essential components needed for GPU kernel execution. For example, device management APIs enable the selection of a specific GPU for executing GPU code, while execution control APIs facilitate launching GPU kernels on the chosen GPU. These APIs are typically encapsulated within user-space GPU libraries, which interact with the GPU kernel driver to perform their functions.

\subsection{GPU Compilation}
A specialized compiler is utilized to process GPU-enabled source code. It first pre-processes directives, macros, and includes. It also splits the code into parts that will run on the device (i.e., the GPU) and those that will execute on the host (i.e., the CPU). Host code is compiled using standard compilers such as gcc or clang. GPU code is lowered into an intermediate representation suitable for further compilation. HIP for example uses LLVM IR. During this process, any optimization passes are carried out on the compiled code. If the GPU target architecture is known, the code is lowered down to a format suitable for execution on the GPU. Otherwise, the intermediate representation is stored in the combined host and GPU binary.

\begin{figure}[!htb]
  \centering
  \includegraphics[width=0.75\columnwidth]{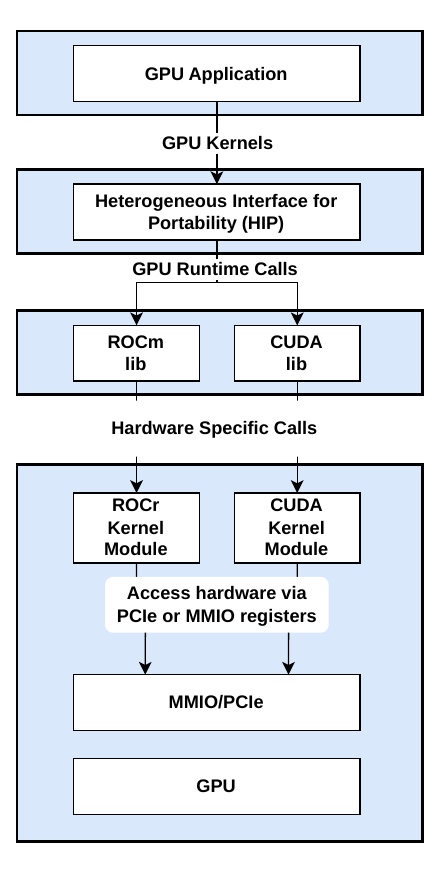}
  \caption{GPU software stack showing a number of abstraction layers that separate the application interface from low level tools that may be hardware-specific. A similar general approach will be employed with working with quantum hardware based accelerators. }
  \label{fig:gpustack}
\end{figure}

\subsection{GPU Scheduling and Execution}
When the program is executed, the operating system loads the binary into
memory and begins executing the host code. Any invocation of the GPU API
results in the execution of the GPU library code, which interacts with the
GPU device driver to perform operations such as GPU memory allocation,
context creation, and device management. When the host code launches a GPU
kernel, the GPU API looks up the function symbol in the kernel symbol table
within the ELF binary. If the GPU function has already been compiled to the
target GPU architecture, it is queued on the GPU scheduler. If it is still
in an intermediate representation (IR) format the GPU API compiles it down to the specific machine code for
the GPU architecture present in the system. This Just-In-Time (JIT)
compilation is optimized for the current hardware and the compiled code is
often cached to avoid recompilation in future executions, significantly
improving performance.

Once compiled, the kernel is queued on the GPU
scheduler, which manages the execution order of multiple kernels to
efficiently utilize the GPU resources. On program completion, the GPU
library performs finalization, ensuring that all GPU operations are
completed, memory is deallocated, and other resources are cleaned up
gracefully. This includes releasing any device contexts, freeing allocated
memory, and ensuring that the GPU is in a clean state for future operations.

\subsection{Discussion}
The GPU software stack described in this section serves as a good template for the QC software stack, refer to Figure~\ref{fig:software-layers}. Similar to the control interfaces shown in Figure~\ref{fig:gpustack}, QC will require a comparable set of interface APIs that provide the appropriate level of abstraction for applications. The GPU compilation process shares many parallels with quantum circuit compilation, where circuits are lowered to an intermediate representation and further optimized into a hardware-specific format, following the JIT compilation process. Finally, circuits will need to be scheduled and executed on the target platform raising similar challenges to host/GPU allocation and coordination. These points will be elaborated on further in Sections~\ref{sec:resmgmt-modes}, \ref{sec:tools-ee}, and~\ref{sec:hpcqc-arch}.

\section{Application Patterns}
\label{sec:app-patterns}

Although hybrid classical-quantum algorithms inherently rely on both types of hardware, it is useful to classify application patterns based on their varying computational demands. Such distinction is important, given the significant asymmetry in the availability of quantum computers in the foreseeable future. In both current, and even next generation of QC/HPC ecosystems, quantum hardware deployment will be limited (i.e., only a single or few quantum devices will likely be available), in contrast to classical compute infrastructure, which will continue to be the dominant computational resource. 
Below, we distinguish between different application patterns based on their need to utilize quantum and classical hardware. These approaches span from those requiring minimal QC resources to implementations where quantum computation dominates the total resource allocation\footnote{Although often the classical HPC compute requirements are expressed in terms of a measure such as \textit{node-hours}, and equivalent can be used for quantum devices, due to the generally very limited availability of the latter, it may initially be more useful to discuss the (quantum especially) requirements in terms of a fraction-of-the-total that are available.}.
While it is expected that all these application types are treated as first-class citizens by the underlying software tools (e.g., ORNL's QFw), explicitly identifying the core differences between them, can help optimize resource allocation and improve overall runtime efficiency in hybrid computing environments.

\begin{enumerate}

    \item \textit{High quantum, low classical resource usage: \label{pattern:highQuantumLowClassical}} 
This category includes applications where the quantum resource requirements dominate the computational workload. Here the classical resources are primarily used for (usually fast, and requiring a minimal HPC node count to execute) pre- and post-processing tasks that may include circuit generation or transpilation, and classical analysis of measurement results. 
   
    \item \textit{Low quantum, high classical resource usage:\label{pattern:lowQuantumHighClassical}} 
        In this category, the core of computation is classical in nature and only a small amount of quantum resources are required. This pattern mainly expands on applications currently encountered in HPC ecosystems, where many HPC nodes are utilized over long times, but where quantum resources may be invoked very selectively, only for specific parts of the total computational task. 

    \item \textit{Roughly equal quantum and classical resource usage:\label{pattern:equalQuantumAndClassical}} 
        Here, quantum and classical computations are more balanced, each playing a substantial role in the total computational task, and neither dominating the total resource requirements. 

\end{enumerate}

Having a framework (and crucially a scheduler) that is able to understand these different application patterns, can help minimize resource idle time, and thus maximize overall hybrid system performance of the full QC/HPC system. 
For example, in the case of applications that adhere to Pattern~\ref{pattern:lowQuantumHighClassical} above, in order to optimally utilize the (highly scarce) quantum hardware, the scheduler may need to interleave quantum execution of multiple hybrid jobs, whereas in the case of jobs that mainly adhere to Pattern~\ref{pattern:highQuantumLowClassical}, the scheduling becomes more straightforward, as jobs can be largely scheduled sequentially. See Section~\ref{sec:resmgmt-modes} for further discussion.

\subsection{Time-Sensitive Classical-Quantum Computation}
\label{sec:timeSensittive}

Some hybrid applications may also have a hard constraint on the latency between interleaved classical and quantum parts of the computation. One example of such a scenario includes quantum circuits with mid-circuit measurements (see, e.g., \cite{Rudinger2022midcircuit}), where parts of the algorithm can be conditionally executed depending on measurement results of a qubit (or many qubits) \textit{during} (i.e., mid-way through) the quantum evolution.  
In the simplest variant of above paradigm, the classical processing involved can be fairly minimal, and in practice can be often done on the close-to-quantum hardware electronics (e.g., on a FPGA- or ASIC-based devices, with minimal communication latency to the quantum hardware). One can, however, also imagine more involved scenarios, where such mid-circuit measurements spawn complicated, even HPC-based computations, which in turn determine what further quantum operations should be executed. This latter operational pattern is currently not practical, due to the very limited coherence times\footnote{In simple terms, a ``coherence time'' of a quantum system defines a timescale beyond which the environmental noise can destroy a quantum state, thus making a quantum computation no longer quantum.} of the available, even state-of-the-art, quantum computers. As quantum hardware evolves, however, and some level of fault-tolerance is achieved, the timescales on which quantum devices will stay coherent will be extended dramatically (to seconds, minutes, or beyond), thus reducing the hard-time constraints on how long hybrid computations (and related communications) that require time-sensitive processing, can take. It is worth pointing out that the inability of a hybrid QC/HPC systems to keep pace, will generally result in a reduction of the fidelity of obtained results (formally, as coherence is gradually lost, quantum states become more mixed, thus introducing classical uncertainty in the qubit measurements, see e.g., \cite{nielsen2010quantum}). 

Another important example where time-sensitive classical-quantum interactions are required is syndrome-decoding (a mechanism by which one can understand how errors can be corrected in a quantum computer, when quantum error correcting codes are being utilized). This process usually involves continually obtaining selective measurements, which then need to be classically processed (i.e., decoded) before future quantum operations are applied, see e.g., \cite{gottesman2010introduction} -- each iteration of this protocol, must happen in a well defined time (that depend on various implementation details such as the hardware being used, or the specific error correction code being utilized). Although researchers previously believed that the decoding process would require substantial classical resources, recent studies~\cite{barber2025real,bausch2024learning} suggest that dedicated FPGA/ASIC hardware may suffice, and HPC-level ecosystems will likely not be directly utilized for syndrome decoding.   
Nevertheless, given the algorithmic usefulness of time-sensitive computations and communications we outline above, it is important for the software stack and tooling to be able to fully support this useful mode of operations.

\subsection{Special Cases and Scheduling Flexibility}

The categorization of hybrid applications based on the relative demand for quantum and classical resources provides a useful starting point for designing resource allocation strategies and guiding scheduler behavior. However, in practice, more fine-grained classifications may also be warranted. These could account for factors such as algorithmic structure, tolerance to execution delays or job duration.

Moreover, certain applications may require exceptions to general scheduling policies. Examples include benchmarking jobs, debugging or testing runs, and workflows tied to time-sensitive experimental campaigns or calibration events. Such cases may not align neatly with the high/low resource usage patterns defined earlier, and instead demand more dynamic treatment within the scheduler. These scenarios can be accommodated by incorporating metadata into job submissions—such as user---defined priority levels or job types---that enable policy overrides or custom handling.

While the current work does not aim to implement these mechanisms directly, we recognize their potential value and consider them an important direction for future extension of this effort.

\section{Resource Management}
\label{sec:resmgmt-modes}

The effective management of resources is an essential component in
a QC/HPC software stack.  A balance must be met between resource
utilization (\emph{platform viewpoint}) and application productivity
(\emph{application viewpoint}).  The platform viewpoint seeks to maximize
utilization of the resources, which keeps the available resources as busy as
possible. The application may have a different objective, whereby the focus
is on minimizing time to solution.  This latter viewpoint may care less
about overall utilization if it simplifies the programming experience or
provides benefits strictly to the singular application (i.e., ignoring
consequences to other users of the shared resource).
These resource management viewpoints are not unique to the QC/HPC context,
but do help guide choices for interfaces and policies to support coordinated
use cases that emerge in coupled QC/HPC systems.

\subsection{Allocation}
\label{sec:resmgmt-modes:allocation}

\begin{figure}[!ht]
  \centering
  \begin{subfigure}[b]{0.49\textwidth}
    \centering
  \includegraphics[width=0.85\columnwidth]{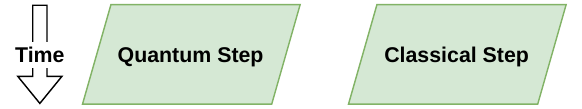}
    \caption{Simultaneous Allocation \medskip}
    \label{fig:alloc:simultaneous}
  \end{subfigure}
  \begin{subfigure}[b]{0.49\textwidth}
    \centering
  \includegraphics[width=0.85\columnwidth]{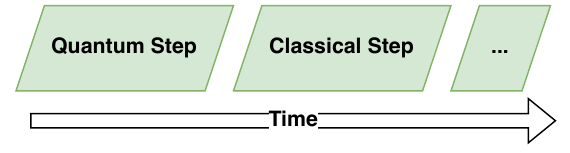}
    \caption{Interleaved Allocation}
    \label{fig:alloc:interleaved}
  \end{subfigure}
  \caption{Allocation strategies to support hybrid QC/HPC applications.}
  \label{fig:alloc}
\end{figure}

The allocation of computational assets is driven by the application usage patterns, which involve quantum and classical steps that determine the allocation strategies (Figure~\ref{fig:alloc}). The application patterns outlined in Section~\ref{sec:app-patterns} identify three general usage modes, characterized by the application's balance of quantum and classical computing (i.e., \emph{High QC/Low classical}, \emph{Low QC/High classical}, and \emph{roughly equal}).

When quantum and classical computing power must be allocated concurrently and remain reserved for the same duration, we refer to this as \emph{simultaneous allocations} (Figure~\ref{fig:alloc:simultaneous}). An \emph{interleaved allocation} allows for independent reservations, potentially overlapping or forming a chained sequence (see Figure~\ref{fig:alloc:interleaved}). For example, HPC capabilities might be used first for pre-processing, followed by quantum systems for execution, with results stored for later post-processing on HPC infrastructure. We use ``HPC" instead of simply ``classical" in this example to emphasize the allocation dynamics of hybrid QC/HPC applications.

\begin{figure}[!t]
  \centering
  \includegraphics[width=0.85\columnwidth]{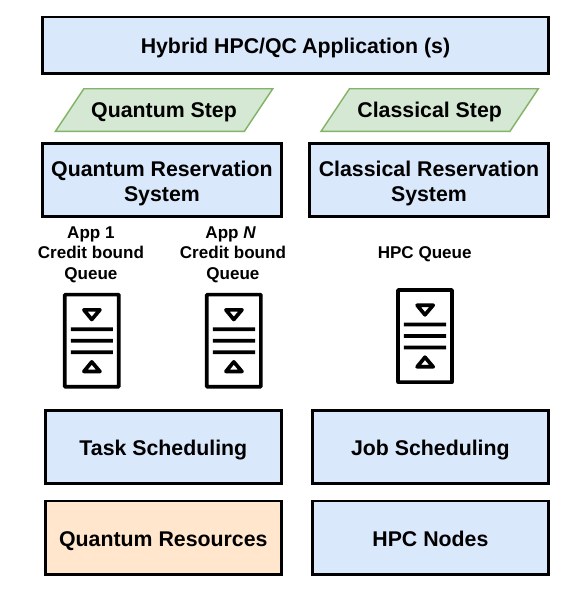}
  \caption{Illustration of hybrid QC/HPC application resource management.
    The classical reservation system uses a system wide queue.
    The quantum resources are exposed via a reservation system that
    determines the scheduling of tasks (ciruits).  The illustration
    highlights our envisioned approach to the quantum resource management
    whereby credit-based application queues help provide practical
    time bounds needed for hybrid applications.}
  \label{fig:qc-resmgmt}
\end{figure}

The steps in the hybrid application correspond to work that must be managed
by the quantum and classical reservation systems,
 as shown in Figure~\ref{fig:qc-resmgmt}.  The typical HPC approach has a single work
queue that is used to submit jobs for execution on the HPC compute nodes.
The HPC scheduler selects the appropriate time and location for the job and
grants exclusive access to a set of compute nodes for a fixed time period.
This access is governed by policies that help balance system usage.
Similar to the HPC scenario, the individual
quantum computational devices are used exclusively for the duration
of a given computation (i.e., circuit execution) with appropriately defined policies.

\subsection{Scheduling}
\label{sec:resmgmt-modes:scheduling}

Currently, we are exploring the use of the existing HPC workload manager,
SLURM, to manage both HPC and Quantum resources.  We
use the heterogeneous job (hetjob) capability in SLURM to enable different
resource specifications to be given for the HPC and QC (job) portions of a
hybrid application.  This allows us to support the simultaneous allocation
mode.

We are currently investigating an approach that would leverage
SLURM's features to define generic resources that would represent a QC device.
For example, the Generic RESource~\cite{schedmd:slurm:gres} functionality allows for defining a named resource and quantity that can be used to track availability of quantum resources.
This will allow us to support
both the simultaneous and interleaved allocation strategies. This integration of both HPC and QC
devices under a single resource management system will enable more detailed
investigations into different scheduling policies and prioritization schemes.

\subsection{Multi-resource Scheduling}
\label{sec:resmgmt-modes:multi-scheduling}

Hybrid QC/HPC applications 
leverage multiple resources to solve problems
that are computationally expensive. The resource requirements are provided
to help guide scheduling decisions for allocating access to the computing
devices.  As mentioned previously, the traditional HPC approach is to block
allocate a set of compute nodes for exclusive (single user) access.  Whether
using a simultaneous or interleaved approach, the QC resources must be
usable during the time window that aligns with the HPC compute allocation.  This
introduces timing constraints for the hybrid applications, which involve
coordination among different scheduling and resource management systems.

The quantum resource manager is driven by requests to execute tasks (i.e.,
quantum circuits).  
The system tracks what quantum devices are in-use, and
allocates available devices to run tasks (quantum circuits).  The tasks run
exclusively on the quantum devices.  The allocation of the quantum devices
is governed entirely by the resource manager to maximize utilization of the
scarce quantum resources.

The coordinated scheduling of the HPC and QC resources introduces several
interesting problems. 
Chief among these is the fact that task duration
is not uniform across the HPC and QC resources. In the case of a
simultaneous allocation, the HPC portion uses the QC hardware and must have
some quality of service on quantum task activation (i.e., bounded request
time). Otherwise, the HPC resource will idle waiting for results from the QC
resource.  Therefore some form of ``\emph{soft allocation}'' is needed to
constrain the activation time for quantum task submissions.  A credit
system can be used to provide these service agreements in order to meet
general bounds on execution.

The QC/HPC software stack needs to provide support for this two-level
scheduling problem to manage quality of service constraints for the hybrid
applications using combined resources.  The current plans are to
explore the creation of dynamic work queues for the quantum application
contexts, as illustrated in Figure~\ref{fig:qc-resmgmt} (e.g., \emph{App N
credit bound queue}).  These hybrid application queues would help control
time bounding via a credit system to help throttle requests to the quantum
resource.  This is similar to other techniques employed in HPC where shared
resources need to be used during phases of the application and need some
degree of quality of service guarantees (e.g., network latency/bandwidth, parallel
filesystems).  The credit-based approach is often useful to ensure
provisioning of the shared resource. The hybrid application requires
access to quantum resources that is managed via a ``\emph{soft
reservation}'' to ensure bound on quantum task activation time. This will require
development of new mechanisms and scheduling capabilities to expose quantum
resource load to enable more advanced coordinated scheduling of hybrid applications.

\section{Hybrid QC/HPC Application Preparation}
\label{sec:tools-ee}

\begin{figure}[!tb]
  \centering
  \includegraphics[width=0.8\columnwidth]{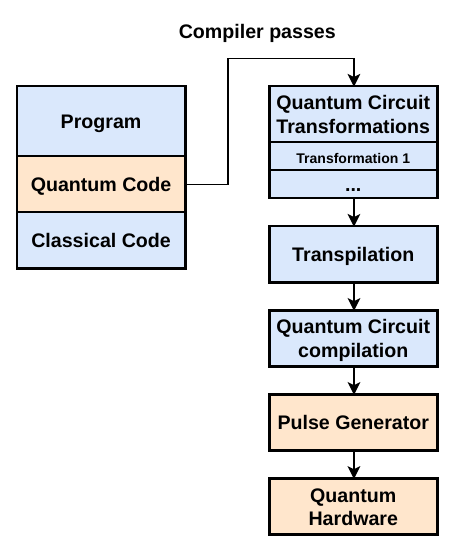}
  \caption{Quantum circuit processing passes.}
  \label{fig:QCircPhases}
\end{figure}

Having examined various application types and resource management approaches for hybrid quantum-classical computing, we now turn our attention to the critical preprocessing steps needed to prepare these hybrid applications for successful execution. A hybrid application consists of two segments: classical and quantum code. The classical code follows standard handling procedures, while the quantum code undergoes several compiler passes before reaching the hardware, as illustrated in Figure~\ref{fig:QCircPhases}. Typically, a quantum circuit is programmed using generic gates, then lowered to an intermediate representation (IR). User-specified transformations, such as circuit cutting and gate reduction, are applied to optimize the circuit. The circuit is then transpiled into hardware-compatible gates, which are ultimately converted into pulses by the quantum controller for execution. 

Hybrid applications can be either interpreted or compiled, requiring a unified software stack to support both approaches. In the interpreted case, all outlined steps are executed on demand once the application builds and runs the circuit, with classical processing up to the pulse generation stage. The proposed software stack abstracts the steps beyond circuit construction while providing configurable controls for quantum circuit transformations, aligning with familiar frameworks like Qiskit \cite{qiskit2024}.

\begin{figure}[!h]
  \centering
  \includegraphics[width=0.8\columnwidth]{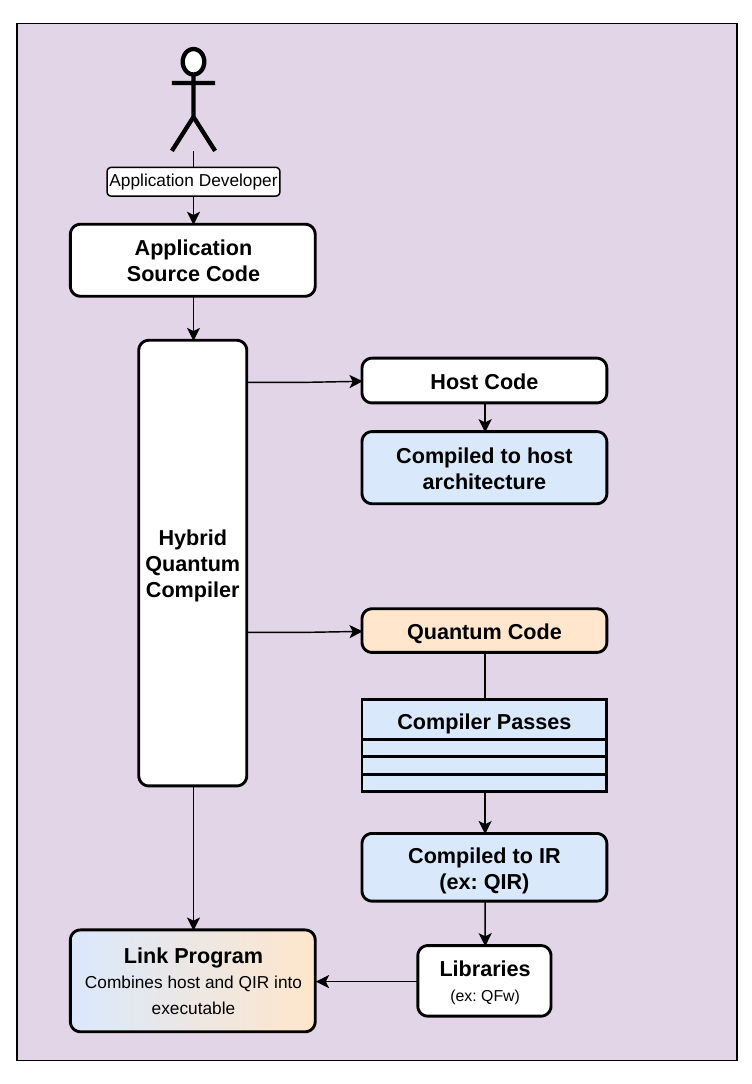}
  \caption{Quantum compiler toolchain.}
  \label{fig:compile}
\end{figure}

The compiled case differs in that some of the quantum processing passes can occur while compiling the hybrid application. The GPU compilation process described in Section~\ref{sec:back-traditional} offers a valuable framework that can serve as a model for developing the analogous process in QC. Figure~\ref{fig:compile} illustrates a high-level view of a quantum enabled compilation flow.

A hybrid application delineates quantum tasks using specialized syntax. A quantum-enabled compiler is then able to pre-process and split the code into host and quantum segments. The quantum segments are lowered into an IR, where compiler passes optimize the quantum operations further. These compiler passes can be configured by the user. If the target quantum hardware is known, the circuit can be lowered further into hardware specific IR; otherwise, it remains in the generic IR form.

The host code 
is compiled into the target architecture using standard compilers, e.g., \command{gcc} or \command{clang}. Both the IR and the compiled host code are then linked against the necessary libraries into a binary that can be executed on classical computing resources.

The target quantum hardware may not be known at the time of the hybrid application compilation. Even if it is known, the dynamic nature of quantum circuit formulation and ongoing hardware calibration, which can alter specific hardware attributes, necessitate an additional transpilation pass followed by translation into pulses. The same infrastructure that processes quantum code, from lowering it to a generic IR to generating pulses, can be utilized both during compilation and at runtime when the hybrid application requests a quantum circuit execution. Therefore, it is essential to formalize the interfaces to that infrastructure, henceforth referred to as the Quantum Toolchain, as shown in Figure~\ref{fig:common-interfaces}.

\section{QC/HPC Architecture}
\label{sec:hpcqc-arch}

\subsection{Software Layer View}
\label{subsec:SoftwareLayerView}

\label{sec:tools-ee:sw-layers}
\begin{figure}[!t]
  \centering
  \includegraphics[width=0.95\columnwidth]{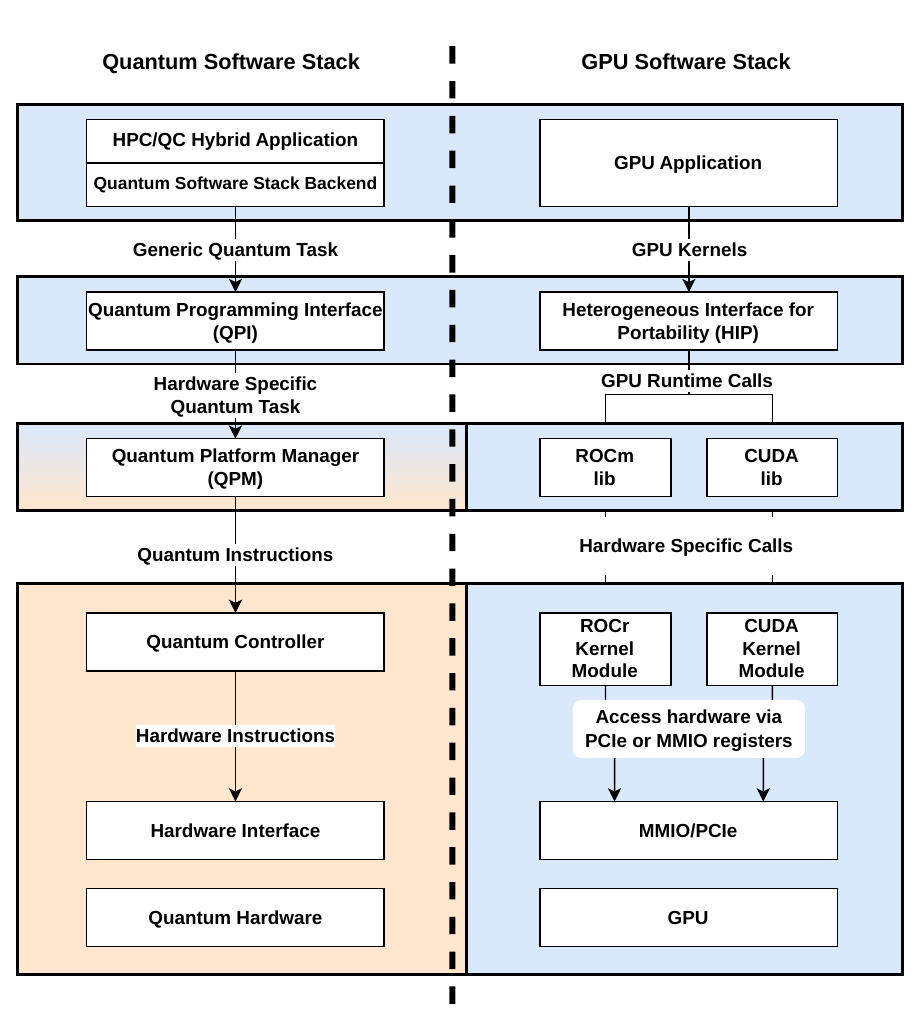}
  \caption{Quantum accelerator software stack layers in contrast to a typical GPU accelerator software stack.}
  \label{fig:software-layers}
\end{figure}

To achieve the described functionality, we propose dividing the software stack into distinct software layers, with each layer providing an appropriate abstraction level to the layer above it as shown in Figure~\ref{fig:software-layers}. In that diagram, we draw parallels between the GPU software stack layers and the proposed quantum software stack layers.

The highest layer of the software stack is the hybrid application itself, which can run either through an interpreter or as a compiled executable. Similar to how an MPI application initializes the MPI library, the hybrid application must initialize the software stack. This initialization process enables the software stack to discover available quantum resources and execute the necessary setup procedures. Once initialized, the application can leverage the most appropriate circuit-building package for its needs. For interpreted applications, Qiskit is one example of such a package.

The proposed software stack provides a backend for a select set of circuit-building packages. The purpose of this backend is to allow the application to use these packages as it normally would, with the only difference being the selection of the backend. This approach enables application developers to build and test their code locally using simulators before seamlessly transitioning to a QC/HPC environment when ready. The backend communicates with the rest of the software stack through the Quantum Programming Interface (QPI) layer. 

The QPI layer offers several categories of APIs designed to simplify interactions with the software stack. These include initialization and finalization of the software stack, device management for discovering and configuring quantum devices, tool management for defining and configuring compilation passes, execution control for launching quantum tasks, resource management for configuring how resources are allocated, and result and error handling. The QPI is implemented as a library linked against the hybrid application and also provides Python bindings to support Python-based applications. The QPI can be accessed directly by the hybrid application or, when used with a package like Qiskit, through the provided backend. 

To submit a quantum task for execution, the application uses the execution control APIs provided by the QPI layer. After processing the task, the QPI layer hands off a hardware-specific quantum task to the Quantum Platform Manager (QPM). The QPM serves as a hardware abstraction layer, offering APIs to facilitate quantum task submission, result retrieval, and queries regarding device status, attributes, and configuration.

The QPM is implemented as a plugin-enabled library that includes a set of utility operations for all hardware providers. The plugin architecture allows each quantum hardware provider to develop their own plugin that implements the QPM API. Common utilities like an RPC communication layer or a scheduling system will be provided by the QPM library. These are intended to ease plugin development. This modular design enables the software stack to support multiple quantum hardware platforms without requiring modifications to the higher layers. 

\subsection{Interface Normalization}

\begin{figure}[!htb]
  \centering
  \includegraphics[width=0.75\columnwidth]{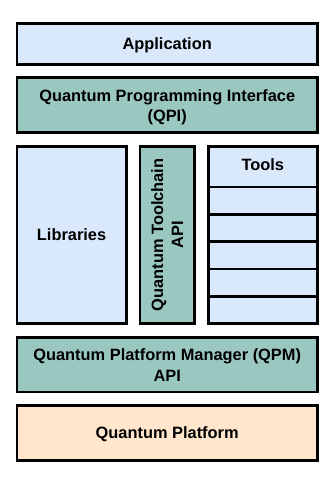}
  \caption{The diagram illustrates the proposed common interfaces that enable interoperability across different implementations. The hybrid application interacts with the software stack through the Quantum Programming Interface. Quantum circuit transformation tools can be integrated via the Quantum Toolchain Interface, while quantum hardware connects to the software stack by implementing the Quantum Platform Manager API.}
  \label{fig:common-interfaces}
\end{figure}

Building on the software layer view introduced in Section~\ref{subsec:SoftwareLayerView}, we further define a set of normalized interfaces, as illustrated in Figure~\ref{fig:common-interfaces}. These interfaces standardize interactions across different hardware and software layers, reducing complexity for hardware vendors and enhancing interoperability. By formalizing these interfaces, different entities can independently develop various components of the software stack while ensuring seamless integration between them. 

Two key interfaces, the Quantum Programming Interface (QPI) and the Quantum Platform Manager (QPM), were mentioned earlier and will be discussed in more details in the following sections. Positioned at the top and bottom of the stack, respectively, their primary role is to standardize interactions between the application layer and the underlying quantum hardware. The first providing an application-friendly interface and the latter a hardware-friendly interface.

\subsection{Quantum Platform Manager}
The Quantum Platform Manager (QPM) API provides a unified abstraction layer for quantum hardware, enabling the software stack to remain hardware-agnostic by exposing hardware capabilities and features consistently. However, given the diverse modalities and varying hardware capabilities, a single generic API risks being overly broad, making it challenging to express hardware configurations and features effectively. To address this, the QPM API is divided into three categories. The first, Resource Management APIs, allows the resource manager to allocate resources for hybrid applications. The second, Runtime APIs, enables hybrid applications to execute quantum tasks during operation. The third category, still under investigation, includes hardware-specific APIs that focus on low-level primitives, such as gate and pulse level control, tailored to unique hardware features. These hardware-specific APIs are designed to map directly to hardware capabilities, while more complex logical operations are handled by higher-level layers, such as the Quantum Programming Interface (QPI). A listing of the QPM APIs is provided in the supplementary material.

\subsection{Quantum Programming Interface}
The Quantum Programming Interface (QPI) acts as a high-level abstraction layer that simplifies how applications interact with diverse quantum hardware. It provides a consistent, application-friendly API that presents a unified view of available resources, allowing applications to configure their execution environment according to their specific needs. This high-level perspective contrasts with the low-level, hardware-oriented API of the Quantum Platform Manager (QPM), which is designed to be closer to kernel driver interfaces and more amenable to vendor implementation. By handling common resource orchestration tasks at the QPI level, the framework avoids redundant implementations across different QPMs. Through the standardization of essential constructs, the QPI promotes consistency across varied quantum hardware platforms, even when individual QPMs expose differing capabilities. Unlike comprehensive quantum programming frameworks such as Qiskit or PennyLane, the QPI does not define a programming paradigm. Instead, it focuses on enabling efficient integration and execution of quantum tasks on heterogeneous backends. At the same time, it supports access to hardware-specific optimizations via QPM-provided APIs, striking a balance between portability and performance.

\sloppy Beyond abstraction, the QPI introduces a mechanism for configuring and managing the quantum software toolchain. Through its tool management APIs, users can define compiler passes, including circuit cutting or gate reduction, and delegate task execution across available high-performance computing (HPC) resources. The QPI is responsible for orchestrating the software stack, selecting appropriate quantum backends, and tailoring the final transpilation stage based on the targeted hardware. It also supports event-driven execution by offering constructs for monitoring task completion and system-level events such as noise fluctuations, calibration updates, or hardware failures, thereby improving operational transparency. To support efficient quantum task execution, the QPI provides a q-stream construct that groups quantum tasks for sequential or parallel execution. These tasks can be explicitly assigned to resources by the user or dynamically scheduled by the QPI based on resource availability. Together, these features position the QPI as a key architectural component for enabling scalable, hybrid quantum-classical workflows.

\subsubsection{QPI Object View}
\begin{figure}[!t]
  \centering
  \includegraphics[width=0.95\columnwidth]{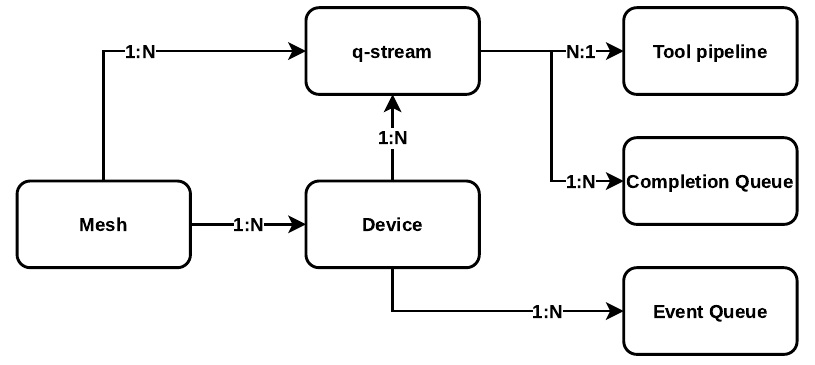}
  \caption{Quantum Programming Interface (QPI) object relationship diagram.}
  \label{fig:qpi-design}
\end{figure}

The QPI design, illustrated in Figure~\ref{fig:qpi-design}, abstracts access to one or more quantum resources through a set of logical constructs. At its core, the Mesh object encapsulates one or more quantum devices, with operations applied to the Mesh affecting all contained devices. These operations are managed by a q-stream object, a container for a sequence of operations—ranging from simple device queries to full quantum task execution, that can be executed together. Multiple q-streams can be bound to a single Mesh, each representing a distinct set of operations configured to run sequentially or in parallel. A Mesh can also distribute operations across its devices, supporting either embarrassingly parallel execution or complex entanglement procedures, such as distributing a quantum circuit across multiple devices within the same Mesh.

Each quantum resource is represented as a device object within a Mesh, and individual devices can also have one or more q-streams bound to them, with operations restricted to the specific device. An event queue object, bound to a device, captures events such as noise incidents, calibration updates, or hardware faults; multiple event queues can be created to handle different event types, but events are ignored if no queues are bound. Similarly, a completion queue, bound to a q-stream, receives notifications when operations in the q-stream complete, such as quantum task results or calibration outcomes. If a q-stream runs operations in parallel, notifications may arrive out of order; otherwise, the QPI ensures ordered delivery. A q-stream can have multiple completion queues, each dedicated to specific completion types.

Finally, the tool pipeline object manages tools applied to quantum tasks, such as compiler passes for circuit cutting or gate reduction. A single tool pipeline can be bound to multiple q-streams, ensuring consistent tool application across all quantum tasks originating from those q-streams. These constructs collectively enable the QPI to streamline interaction with quantum hardware, monitor operation completion and system events, and optimize hybrid quantum-classical workflows. The QPI module is currently under active development. The proposed API is included in the supplementary materials.

\subsection{Quantum Toolchain}

Another critical interface is the Quantum Toolchain API. As discussed in Section~\ref{sec:tools-ee}, quantum tasks require transpilation at runtime, even if they have already undergone compilation at compile-time. This transpilation process is not limited to converting a high-level quantum task into a hardware-compatible form; it also includes essential optimizations such as, gate reduction and circuit cutting. These operations refine quantum tasks to maximize efficiency and adapt them to the constraints of specific hardware backends.

The objective of this software stack proposal is to formalize the interface to these tools, enabling tool developers to extend the stack without requiring modifications to other components. By establishing a well-defined Quantum Toolchain API as in Figure~\ref{fig:common-interfaces}, the stack ensures flexibility, allowing new tools and optimizations to be integrated seamlessly while preserving interoperability with existing software and hardware layers. 

Each tool will take quantum programs expressed in QIR~\cite{qirspec} or OpenQASM~\cite{cross2017openquantumassemblylanguage}, perform circuit transformation, and produce a modified QIR or OpenQASM representation that can be passed to the next tool in the pipeline. This process forms a toolchain of operations, where a quantum task is introduced at the top of the pipeline and progressively refined through a series of transformations. At the end of this pipeline, the final version of the quantum task is fully optimized and lowered into a hardware-compatible format to be accelerated to the corresponding backend~\cite{qireerepo,wong2024qireepreprint,mccaskey2020xacc}. For example, the execution engine for QIR~\cite{qireerepo} (QIR-EE) can be invoked for the purposes of parsing, interpreting and executing QIR across multiple hardware platforms. The software stack, through the QPI, provides applications with the ability to configure this pipeline to suit their specific needs, ensuring flexibility in how quantum tasks are processed and optimized before execution.

Additionally, the tools themselves can leverage HPC resources for execution, as certain transformations, such as large scale circuit optimization or gate reduction, can be computationally intensive and time-consuming. By utilizing the available HPC allocation, these operations can be distributed across multiple compute nodes, significantly reducing processing time and enabling efficient handling of complex quantum tasks.

\subsection{Architectural Overview}
\label{sec:hpcqc-arch:overview}

\begin{figure}[!htb]
  \centering
  \includegraphics[width=0.95\columnwidth]{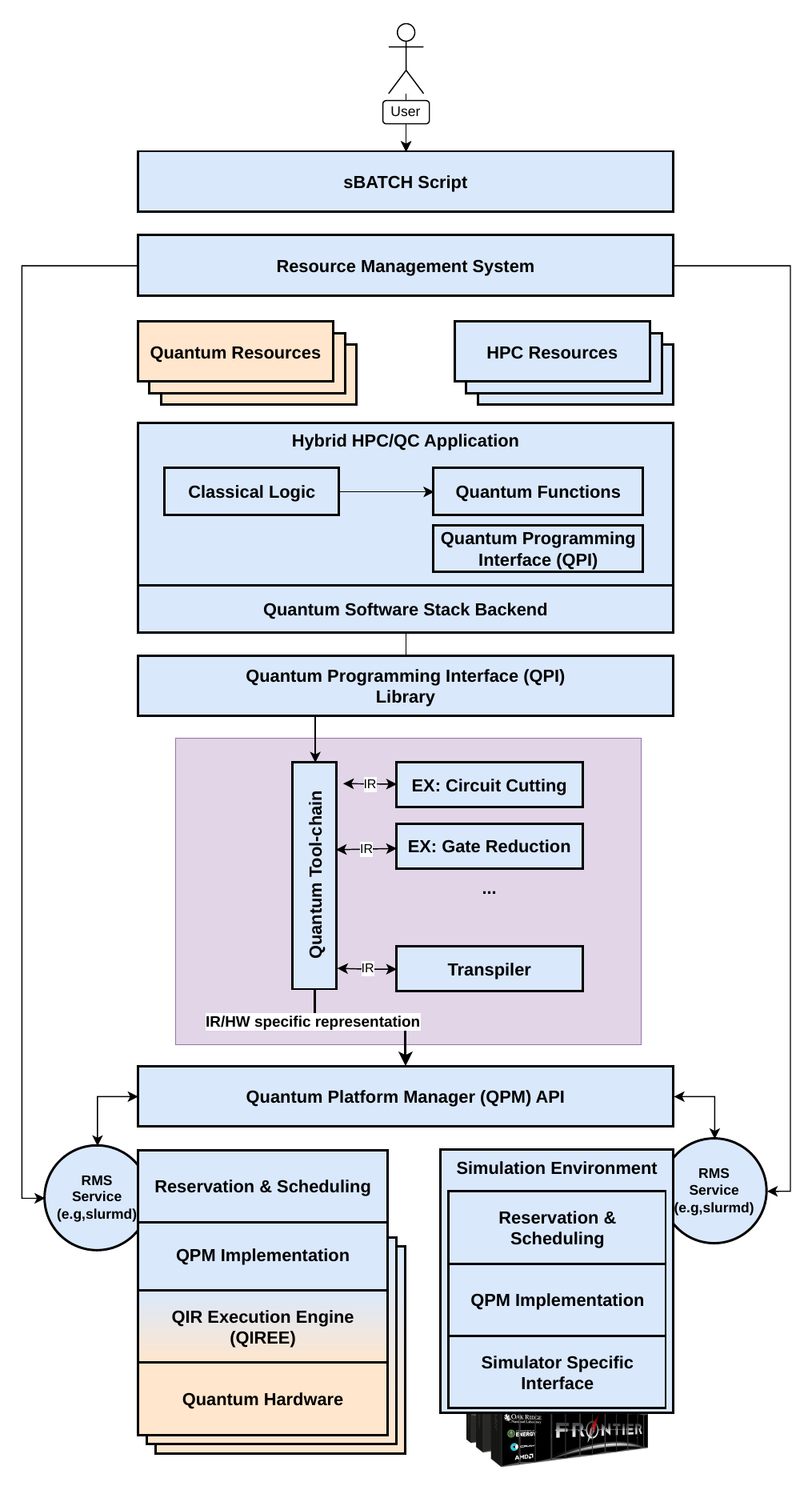}
  \caption{QC/HPC detailed software stack architecture}
  \label{fig:qchpc-arch}
\end{figure}

The architecture depicted in Figure~\ref{fig:qchpc-arch} outlines a more detailed breakdown of the software stack. The Quantum Framework (QFw) includes the quantum software stack backend, the QPI Library implementation, the quantum toolchain implementation and the simulation environment, which includes multiple simulator support, e.g., TNQVM~\cite{Nguyen2021TensorNQ} and NWQ-Sim~\cite{li2021svsim}. It also includes a set of scripts which give the user a familiar HPC like interface. The prototype we have implemented thus far realizes key components of the architecture and serves as a means to validate our conceptual designs. It has been described in greater detail in our previous work~\cite{beck2024quantum} as well as in the supplementary material.

We will analyze the architecture with a top-down approach. Typically, users would compose an sbatch script for submission as a SLURM job on a supercomputer like Frontier, where the script specifies the resources needed for the job, which could be HPC resources, QC resources, or a combination thereof. The QFw provides a set of startup scripts that initialize its infrastructure. These scripts should be incorporated into the sbatch script when launching a job. The design capitalize on two SLURM features, Generic RESource (GRES) and Heterogeneous Job features. The latter is used to identify two sets of resources which need to be allocated at the same time, as shown in Figure~\ref{fig:listing:hetjob}.

\begin{figure}[!htb]
\begin{lstlisting}[
  language=bash,
  showspaces=false,
  columns=fullflexible,
  breaklines=true,
  frame=single,
  emphstyle=\bf,
  basicstyle=\footnotesize\ttfamily\scriptsize,
  stringstyle=\ttfamily,
  breaklines=true,
  numbers=left,
  numberstyle=\tiny,
  stepnumber=1]
#!/bin/bash

# job component 1
#SBATCH -A stf008
#SBATCH -N 10
#SBATCH --partition=compute
#SBATCH -t 1:00:00

#SBATCH hetjob

# Heterogeneous job definition
# for the QC node
#SBATCH --partition=quantum 
#SBATCH --nodes=1
#SBATCH --ntasks=1
#SBATCH --gres=qc:QC:1
#SBATCH --time=01:00:00
\end{lstlisting}
  \caption{Example of a SLURM heterogenous job that includes two parts representing the HPC and QC sides of the hybrid application; 10~nodes allocated for the HPC portion, and 1~node for the QC portion.} 
  \label{fig:listing:hetjob}
\end{figure}

\begin{figure}[!htb]
\begin{lstlisting}[
  language=bash,
  showspaces=false,
  columns=fullflexible,
  breaklines=true,
  frame=single,
  emphstyle=\bf,
  basicstyle=\footnotesize\ttfamily\scriptsize,
  stringstyle=\ttfamily,
  breaklines=true,
  numbers=left,
  numberstyle=\tiny,
  stepnumber=1]
# Setup the QFw infrastructure
qfw_setup.sh
# Run the application
qfw_srun.sh python3 application.py
# Teardown the QFw infrastructure
qfw_teardown.sh
\end{lstlisting}
  \caption{Example of QFw helper scripts for framework setup, application execution wrapper and framework teardown.}
  \label{fig:listing:qft-setup}
\end{figure}

The first component indicates a request for ten nodes from the compute cluster while the second component requests one QC node. GRES allows the specification of criteria which is passed down to the GRES plugin aiding in the selection of the quantum resource. As described in Section~\ref{sec:resmgmt-modes}, during this step the QC reservation system is engaged to determine if the QC resource can accommodate the job. Only when the QC and HPC reservation is successful will the hybrid application reservation be granted and the rest of the script execute. The sbatch script can then use the provided scripts to setup the QFw infrastructure, run the application and teardown the QFw infrastructure as shown in Figure~\ref{fig:listing:qft-setup}.

As discussed in Section~\ref{sec:resmgmt-modes}, ensuring the most efficient use of quantum hardware is paramount. Given the high demand and limited availability of QPUs, an effective resource management strategy is essential to maximize utilization and minimize idle time. To address this, the QFw will provide a reservation and scheduling library. This library will be responsible for handling job reservations, ensuring that quantum resources are allocated efficiently. During job reservation, this library will be used to request and secure access to the quantum hardware before the job runs.

The cluster SLURM controller will interact with SLURMd to determine the availability of quantum resources. As illustrated in the corresponding diagram, the reservation library is designed to integrate with the SLURMd plugin running on the quantum gateway via a GRES plugin. The reservation library will leverage the QPM API to query quantum hardware for resource-specific details, such as qubit availability, calibration status, and expected circuit runtime constraints. Based on these criteria, the library will make an informed reservation decision, ensuring that job scheduling aligns with the current state of the quantum hardware. Once a reservation decision is made, the result is communicated back to the SLURM controller, which finalizes the scheduling process and allocates the necessary resources accordingly.

Once the resources are granted, the application begins execution on the classical compute portion of the allocation. This may include HPC nodes or, in cases where only quantum resources are requested, the quantum gateway. The application first calls the initialization routine of the QPI, setting up the necessary components before proceeding with its computational tasks. During execution, the application configures the tool pipeline and launches quantum tasks through appropriate QPI calls. The QPI library ensures that each quantum task is processed through the tool pipeline and ultimately lowered down to the target hardware. To achieve this, the QPI interacts with the QPM API, querying hardware-specific details and selecting the appropriate transpiler based on the returned information. Once the quantum task is processed through the toolchain, the QPI then invokes the QPM APIs responsible for executing the quantum task on the hardware. Since the hardware could be managing multiple jobs, the QPM utility layer, provides a scheduling library, as outlined in the system architecture, to coordinate between the different jobs ensuring tasks are completed within a time bound as discussed in Section~\ref{sec:resmgmt-modes}.

Once execution is complete, the results are propagated back up the stack to the application. From the application’s perspective, this process appears as a standard synchronous or asynchronous function call, abstracting the complexities of quantum execution. Finally, after completing all quantum and classical computations, the application calls the QPI finalization routines, which gracefully clean up allocated resources and terminate the session.

\subsection{Simulation Environment}

Quantum computers are expected to remain a scarce resource for the foreseeable future. As a result, classical quantum simulators will continue to play a crucial role in aiding researchers with testing and debugging their applications before deploying them on actual quantum hardware. To support this need, the QFw infrastructure provides a simulation environment designed to simplify the integration and management of multiple simulator backends.

The simulation environment can be assigned multiple HPC nodes, allowing it to fit seamlessly into the overall software architecture described thus far. It achieves this integration by implementing a QPM plugin, similar to how real quantum hardware interfaces with the software stack. Since each simulator may have unique characteristics and requirements, it is possible for each simulator backend to have its own QPM implementation, ensuring compatibility with the broader quantum execution framework.

Managing the various simulator instances is a key function of the simulation environment. This is accomplished using the PMIx Reference Run-Time Environment (PRTE), a well-established library designed to handle parallel process execution on HPC clusters. A single PRTE instance runs within the simulation environment and manages all allocated nodes. This setup allows applications to utilize multiple types of quantum simulators simultaneously, with PRTE orchestrating their execution.

Since the simulation environment spans multiple HPC nodes, it can distribute quantum task simulations across available resources by launching multiple simulator instances concurrently. If a simulator supports MPI operations, the environment can take advantage of all allocated nodes to run a single large-scale simulation, significantly enhancing circuit simulation capabilities. However, the scalability of this approach is constrained by factors such as the simulator type, the number of available nodes, and the total memory capacity.

By providing a flexible and scalable simulation environment, the QFw infrastructure facilitates experimentation, benchmarking, and debugging. Researchers can test their algorithms in a stable environment before deploying them on quantum hardware, improving efficiency and reducing the need for costly quantum computing time.

\section{Architecture Validation}
\label{sec:validation}

\begin{figure}[!t]
  \centering
  \includegraphics[width=0.95\columnwidth]{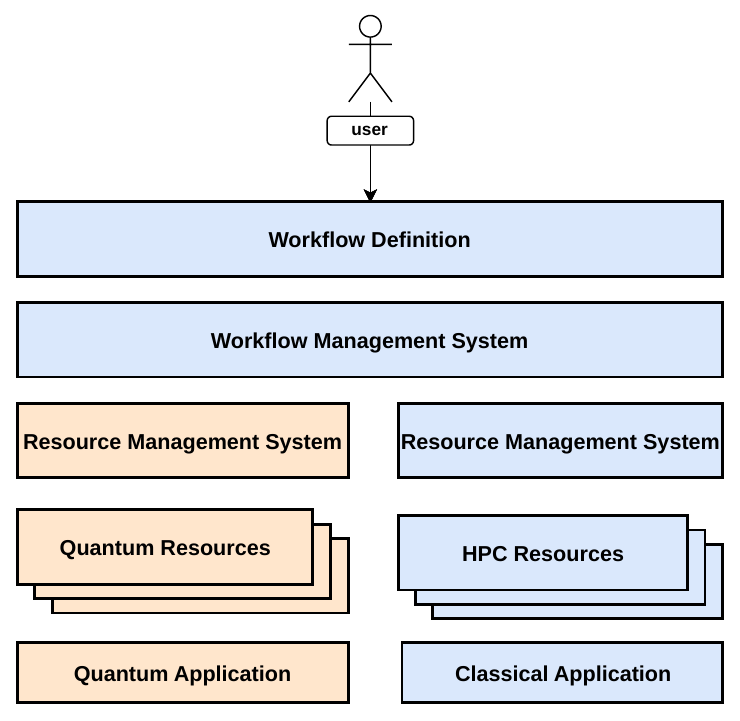}
  \caption{Illustrate the integration of the quantum software stack described in this paper with workflow management systems}
  \label{fig:workflow}
\end{figure}

\subsection{Workflow Integration and Service Management Architecture}

The integration of quantum computing within HPC environments requires robust workflow management and service infrastructure to support diverse operational patterns. Our architecture addresses this through three complementary components: (1)~distributed workflow management, (2)~secure service integration, and (3)~comprehensive system telemetry. The software stack architecture proposed in this paper is designed to work seamlessly with workflow orchestration frameworks that manage distributed quantum-classical workflows across heterogeneous resources.

Traditional HPC ecosystems typically rely on direct SSH access, batch schedulers like SLURM, and manual job submission processes. However, quantum-HPC integration requires more sophisticated infrastructure capabilities including: (1)~programmatic API access for automated systems, (2)~real-time data streaming between experimental facilities and compute resources, (3)~fine-grained authorization and policy enforcement, (4)~workflow orchestration across heterogeneous resources, and (5)~secure communication channels for sensitive quantum algorithms. A service mesh architecture addresses these requirements by providing an infrastructure layer that facilitates secure and efficient communication between distributed services while abstracting networking complexity and enforcing authentication, authorization, and traffic management policies. 

The proposed software stack provides key abstractions for resource allocation, 
and task decomposition, 
enabling efficient coordination between quantum and classical computing elements. As mentioned in Section~\ref{sec:resmgmt-modes}, it will also ensure quantum resources are utilized optimally, by supporting dynamic resource allocation and intelligent task scheduling. These features can be further leveraged by workflow management systems such as Pilot-Quantum~\cite{mantha2024pilot} to handle complex dependency patterns such as variational quantum algorithms and circuit cutting workflows.

OLCF's Secure Scientific Service Mesh (S3M)~\cite{oral2024olcf} will provide the foundational infrastructure for secure and flexible integration of quantum resources within the broader HPC ecosystem. S3M will enable controlled access to quantum and classical resources through policy-driven interfaces, supporting both traditional HPC workloads and emerging quantum-classical hybrid applications. The framework implements sophisticated access control, rate limiting, and centralized authentication, while providing a secure bridge between external workflow systems and internal HPC resources. This architecture will allow quantum applications to leverage HPC resources as computational accelerators within larger scientific workflows, while maintaining operational security and trust boundaries.

Other HPC centers can implement similar capabilities through various approaches:
\begin{enumerate}
    \item \emph{API Gateway Solutions:} Centers using traditional batch schedulers can implement RESTful API layers that provide programmatic access to SLURM/PBS systems while maintaining existing authentication mechanisms.
    \item \emph{Container Orchestration Platforms:} Sites with Kubernetes or OpenShift deployments can leverage service mesh technologies to provide similar traffic management, security, and observability features.
    \item \emph{Workflow Management Integration:} Existing workflow systems can be extended with quantum-aware scheduling and resource management capabilities.
    \item \emph{Cloud-Native Approaches:} Centers adopting cloud-native architectures can utilize managed services from cloud service providers to achieve comparable functionality.
\end{enumerate}
The key architectural requirements (secure API access, real-time streaming workflow orchestration, and policy reinforcement) can be satisfied through different technological implementations while maintaining the same functional capabilities for quantum-HPC integration.

The system also incorporates comprehensive telemetry capabilities that collect operational metrics across both quantum and classical resources. A dedicated telemetry service utilizes the QPM APIs to gather quantum hardware-specific data, including qubit calibration status, error rates, and queue depths. This data is stored in a persistent database alongside classical system metrics, enabling correlation analysis and performance optimization. The telemetry infrastructure supports both real-time monitoring for operational decision-making and historical analysis for resource utilization optimization. Through integration with S3M, the telemetry data could be securely accessed by authorized workflow systems and analysis tools, facilitating informed scheduling decisions and hardware-aware optimization of quantum-classical workflows. 

\subsection{Application Experience}

We tested the QFw framework with an NWQ-Sim backend to tackle a hybrid quantum-classical application: the variational quantum linear solver (VQLS)~\cite{bravo2023variational}. VQLS is a variational approach for solving linear systems of equations of the form $A\mathbf{x}=\mathbf{b}$, offering a scalable solution for quantum linear solvers in existing NISQ devices. It has practical applications in various fields, including fluid dynamics~\cite{demirdjian2022variational,ye2024hybrid}, which is our specific focus--solving canonical fluid flow problems using NISQ devices \cite{gopalakrishnan2024solving,meena2024towards}.

The algorithm formulates an optimization problem where a quantum circuit represents the solution vector $\mathbf{x}$, denoted as $\hat{\mathbf{x}}$. The objective is to minimize the error in reconstructing $\mathbf{b}$ using $\hat{\mathbf{x}}$, which defines the cost function to be minimized. The optimization is performed iteratively (over multiple epochs), with a classical optimization technique adjusting the parameters of the quantum circuit at each epoch. The cost function is computed using a QPU. 

We implemented VQLS using Pennylane \cite{bergholm2018pennylane}. The integration with QFw is seamless from the application user's perspective--the only modification required is defining the Pennylane {\tt device} variable, as shown in Figure~\ref{fig:listing:vqls}. For the specific test that we performed, we solved a $2 \times 2$ matrix problem, resulting in a circuit with 3 qubits. We used a simple ansatz involving Hadamard and RY rotation gates on each qubit, giving 3 parameters to be optimized. One function evaluation of this problem for optimization requires execution of 72 circuits.

\begin{figure}[!t]
    \centering
    \begin{subfigure}[b]{0.49\textwidth}
        \centering
    \begin{lstlisting}[language=Python,
                  basicstyle=\ttfamily\small,
                  breaklines=true,
                  breakindent=10pt,
                  postbreak=\mbox{\textcolor{black}{$\hookrightarrow$}}]
# Initial implementation
dev_mu = qml.device("lightning.qubit", 
        wires=tot_qubits, shots=n_shots)
    \end{lstlisting}
    \caption{Initial implementation}
    \label{fig:listing:vqls:initial}
    \end{subfigure}
    \vspace{5pt}
    
    \begin{subfigure}[b]{0.49\textwidth}
        \centering
    \begin{lstlisting}[language=Python,
                  basicstyle=\ttfamily\small,
                  breaklines=true,
                  breakindent=10pt,
                  postbreak=\mbox{\textcolor{black}{$\hookrightarrow$}}]
# QFw porting
from qfw_qiskit import QFWBackend, QFwBackendType, QFwBackendCapability

nwqsim = QFWBackend(
    betype=QFwBackendType.QFW_TYPE_NWQSIM, 
    capability=QFwBackendCapability.QFW_CAP_STATEVECTOR)

dev_mu = qml.device("qiskit.remote", 
        wires=tot_qubits, backend=nwqsim, shots=n_shots)
    \end{lstlisting}
    \caption{QFw port}
    \label{fig:listing:vqls:qfw-port}
    \end{subfigure}
    \caption{Listing showing required changes for port of VQLS using Xanadu's Lightning simulator~\cite{LightningSimulator} to the QFw backend.}
    \label{fig:listing:vqls}
\end{figure}

As we continue exploring different ansatzes for our application, porting our VQLS implementation to the QFw revealed challenges related to optimizer selection. The initial PennyLane implementation was validated using local simulators to assess both accuracy and computational speed. The QFw framework provided easier access to more detailed insights, including a log of executed circuits, queuing times before execution, execution times per circuit, and other performance metrics. 
For example, the total time per epoch for our use case was 342 sec, with an average time of $\approx 0.4$ sec per circuit evaluation. Using the Lightning simulator, running on a local node, the time per circuit evaluation was $\approx 0.003$ sec.
This profiling information revealed that a large number of circuits were being submitted to optimize the circuit parameters at each epoch - 480 circuit evaluations per epoch. This stemmed from the use of a gradient-based optimization method that requires large number of circuit evaluations per parameter update. Initially, we hypothesized that due to the complexity of our end application, a gradient-based strategy, such as stochastic gradient descent, might be necessary to traverse the optimization landscape.

To address this, we replaced the optimizer with a gradient-free optimization approach: the Constrained Optimization by Linear Approximation (COBYLA) method \cite{powell1994direct,pellow2021comparison}, which performs only 1 function evaluation per epoch. This modification reduced the optimization time for our application due to lower number of function (circuit) evaluations - only 1 function evaluation per epoch, resulting in 72 circuit evaluations per epoch. Gradient-based optimization can be useful to tackle certain scenarios with complex function evaluations to compute the gradients. The analysis using QFw denotes that the choice and applicability of the optimizer for our application warrants further exploration. The original Pennylane implementation did not reveal this inefficiency, as local simulators have negligible overhead in circuit execution and parameter updates for such small test cases. Similarly, running on a real QPU through cloud services need not expose this bottleneck either, since the connection latency to the quantum cloud service would overshadow the computational overhead of evaluating numerous circuits. The profiling and debugging capabilities enabled by QFw for simulators can be invaluable for identifying such inefficiencies in hybrid applications.

Looking toward the near future, quantum computers will likely remain standalone machines interfaced with HPC systems via high-speed networks, adding layers of complexity to hybrid quantum-classical workflows. These systems will incur overheads from the surrounding infrastructure---such as latency from framework layers, communication delays between HPC and quantum subsystems, queuing delays, or device---specific I/O bottlenecks. The software stack can often mitigate framework latency by maximizing the use of HPC resources for processing quantum tasks and optimizing network communications. However, these optimizations will not eliminate latency limitations entirely.

Applications will also need to be mindful of circuit generation and execution. As described above, applications like VQLS---or other variational algorithms—when scaled to larger problem sizes, may generate thousands of quantum circuits. Moreover, some of this circuit generation may be inherently sequential, as the structure of a given circuit can depend on results obtained from previously executed ones, thereby limiting opportunities for parallelism and making the overall workflow more sensitive to latency-related overheads.

Nevertheless, efforts should be made to parallelize circuit generation whenever possible. For example, Lu et al. \cite{lu2025lugo} demonstrate a workflow (not involving hybrid Q-HPC execution) that enables parallel circuit generation for the Quantum Phase Estimation (QPE) algorithm. More broadly, application developers should remain mindful of the structure and dependencies within their workflows. Latency constraints that arise from sequential circuit generation or tight classical-quantum feedback loops are often difficult to mitigate implicitly at the framework level. Designing algorithms and workflows that expose opportunities for concurrency, where possible, can significantly reduce end-to-end execution time in hybrid settings.

\section{Related Work}
\label{sec:related}

The space exploring QC integration with existing HPC resources is relatively nascent, with a flurry of activity and collaborations surfacing only within the last few years. Some examples include conceptual renderings of QHPC middleware~\cite{saurabh2023qhpcmiddleware}, an effort at ORNL~\cite{monilwong2025qiris} integrating quantum runtimes in parallel with other accelerators onto CPUs and GPUs within an established task-based kernel framework IRIS~\cite{kim2021iris}, and the Munich Quantum Valley~\cite{hpcqc}, a consortium of research institutions and universities in Bavaria focused on a unified software stack for the HPC-QC ecosystem~\cite{mqv, kaya2024software}. However, other international teaming underscores the importance of pursuing robust integration efforts and and providing frameworks for effectively enhancing conventional computing. For example, HPC-QS~\cite{hpcqs}, funded by the European Union, has intentions to integrate and couple two quantum simulators, with two existing European Tier-0 supercomputers, and to deploy an open European federated hybrid infrastructure. More recently, IBM joined forces with the Riken Quantum Computing Center~\cite{ibmriken}, where the IBM system powered by a 133-qubit IBM Quantum Heron processor, would be co-located and integrated with the Fugaku supercomputer. Additionally, with Pasqal~\cite{ibmpasqal}, they aim to develop a unified programming model built on Qiskit, aiming to integrate quantum and classical computing resources for HPC workflows. Lastly, IQM have partnered with Hewlett Packard Enterprise (HPE) for their version of quantum-HPC integration~\cite{iqmhpe}. Other industry developments that can potentially enable effective hybrid QC-HPC include Nvidia's Cuda-Q~\cite{cudaq} and Riverlane's Deltaflow~\cite{deltaflow}.
\section{Conclusion}
\label{sec:conclusion}

This paper has developed a framework for the integration of Quantum Computing
(QC) with High-Performance Computing (HPC) architectures. Building upon previous work that 
integrated GPU acceleration hardware into HPC, the paper details the
design and functionality of a multi-layered software stack that aims to bridge
these two computational paradigms in a seamless and hardware agnostic way. 
Our approach is to maintain the
familiar HPC user experience while introducing quantum acceleration capabilities,
addressing a range of application motifs that includes 
simultaneous and interleaved QC/HPC workflows.

Our framework emphasizes the necessity for low-latency interaction between classical
and quantum resources. We integrate quantum hardware
with classical HPC by developing standardized APIs such as the Quantum Platform
Manager (QPM) that manages resource utilization effectively. At the top of the stack, 
the Quantum Programming Interface (QPI) abstracts the quantum resources, providing 
a unified interface for applications.  The design
accommodates resource sharing, batch processing of quantum tasks, and the
use of simulation backends for testing and development, all of which
contribute to maximizing the utilization of scarce quantum resources.

The introduction of a Workflow Management System for the QC/HPC interleaved
workflows allows for optimized resource allocation and usage, ensuring that
neither quantum nor classical resources remain idle. Moreover, we
highlight the importance of telemetry and result persistence through a
database system, which not only aids in machine health monitoring but also
enhances the utility of results in both internal and external workflow
contexts, particularly with OLCF's Secure Scientific Service Mesh (S3M).

The methodologies presented in this paper provide a blueprint for
integrating quantum computing into existing HPC infrastructures. They
address critical challenges such as latency, resource management, and
workflow optimization, paving the way for a future where quantum computing
can be seamlessly leveraged to accelerate scientific and computational
tasks. As quantum technologies evolve, these frameworks will need to adapt,
but the foundational concepts introduced here offer a robust starting point
for this integration, promising to unlock new computational possibilities in
scientific research and beyond.

The strategies outlined here begin to expose the complexities involved 
in advancing an adaptive ecosystem model for the next generation of HPC. 
Extensive progress has 
already been made during the transition from CPU-based to GPU-based parallel 
supercomputing over the last two decades, 
culminating in the recent crossing of the exascale barrier (see Section 3). 
The application-dependent challenges presented by integrating quantum resources, however, are
apparent in the detailed development of the quantum framework (QFw) outlined above.  The goals of 
balancing the demands of resource allocation/management, scheduling, data movement between heterogeneous 
processors, and the development of complex workflows that optimally coordinate the multi-user computing 
environment in a hardware-agnostic way present a computing grand challenge for the next decade. The 
long-term aim for the convergence of modeling and simulation, artificial intelligence, and quantum computing 
into powerful and adaptive tools that significantly accelerate scientific discovery is certainly a worthy goal for
the computational science community. 

\section*{Acknowledgments}
This research used resources of the Oak Ridge Leadership Computing Facility at the Oak Ridge National Laboratory, which is supported by the Office of Science of the U.S. Department of Energy under Contract No. DE-AC05-00OR22725. Research sponsored by the Laboratory Directed Research and Development Program of Oak Ridge National Laboratory, managed by UT-Battelle, LLC, for the US Department of Energy.

We thank Dr. Chao Lu (ORNL) for his contributions to the development of the VQLS application, which was used as a representative example in this work.

\bibliographystyle{elsarticle-num}
\balance     
\bibliography{references}


\end{document}